\documentstyle[prl,aps]{revtex}
\twocolumn
\begin{document}
\author{Jian-Qi Shen \\{\it E-mail address}: jqshen@coer.zju.edu.cn}
\address{
Zhejiang Institute of Modern Physics and Department of Physics,
Zhejiang University, Hangzhou 310027, P. R. China}
\date{\today}
\title{Neutron-gravity interferometry experiment: testing Earth's rotating frequency fluctuations via neutron Berry's phases due to spin-rotation couplings}
\maketitle
\begin{abstract}
 ${\mathcal ABSTRACT}$

Neutron spin can be coupled to the Earth's rotating frequency.
Once if the Earth's rotating frequency is {\it time-dependent},
then the neutron will acquire a Berry's topological phase (cyclic
adiabatic geometric phase). So, a potential method to investigate
the Earth's time-varying rotating frequency by measuring the phase
difference between geometric phases of neutron spin polarized
vertically down and up is proposed.
\\
\end{abstract}
In gravity theory, the intrinsic spin of a particle can be
regarded as a {\it gravitomagnetic dipole
moment}\cite{Mashhoon1,Mashhoon2,Shen}. For this reason, it can be
coupled to the gravitomagnetic fields. According to the principle
of equivalence, the rotating frequency of a non-inertial system
can be considered a gravitomagnetic field, and in consequence the
Coriolis force acting upon a moving particle can be regarded as a
Lorentz (gravitomagnetic) force. Thus, a spinning particle in a
rotating frame of reference will be coupled to the rotating
frequency, which is termed the spin-rotation coupling. The
Hamiltonian describing this interaction of particle spin ${\bf S}$
with the rotating frequency $\vec{\omega}$ of a non-inertial frame
is written as $H=\vec{\omega}\cdot{\bf
S}$\cite{Mashhoon1,Mashhoon2,Shen}. The spin-rotation coupling
leads to the inertial effects of the intrinsic spin of a particle,
for instance, although the equivalence principle still holds, the
universality of Galileo's law of freely falling particles is
violated, provided that the spin of falling particle is polarized
vertically up or down in the non-inertial frame\cite{Mashhoon1}.

If the Hamiltonian $H$ of a quantum system is {\it
time-dependent}, then it will give rise to a geometric phase
(Berry's phase)\cite{Berry}. Differing from dynamical phase that
depends on dynamical quantities of systems such as energy,
frequency, velocity as well as coupling coefficients, geometric
phase is {\it independent} of these dynamical quantities. Instead,
it is only related to the geometric nature of the pathway along
which quantum systems evolve. This, therefore, implies that
geometric phase presents the topological and global properties of
quantum systems in time-evolution process, and that it possesses
the physical significance and can thus be applied to various
fields of physics.

If the Earth's rotating frequency is time-dependent, {\it i.e.},
$\vec{\omega}(t)$ whose precession frequency and precessional cone
angle are respectively $\Omega $ and $\theta$, then the
interacting Hamiltonian of neutron spin-rotation coupling is
$H(t)=\vec{\omega}(t)\cdot{\bf S}$. The time-dependent quantum
system is governed by the time-dependent Schr\"{o}dinger equation.
By using Berry's phase formula or Lewis-Riesenfeld invariant
theory\cite{Shen,Shen2}, one can arrive at the expression for
Berry's phase, $\phi^{\rm (Berry)}_{\pm}(T)=\pm\frac{1}{2}\Omega
(1-\cos \theta)$ in one precessional cycle
($T=\frac{2\pi}{\Omega}$), acquired by the neutrons, where the
sign $\pm$ corresponds to the cases of neutron polarized
vertically up and down, respectively.

The variation of the Earth's rotating frequency may be caused by
the motion of interior matter, tidal force and the motion of
atmosphere as well. Once we have information on the Earth's
rotating frequency, it is possible to investigate the motion of
matter on the Earth. Here we suggest a potential approach to
detecting the fluctuations of the Earth's time-dependent rotation:
specifically, measuring the geometric phase arising from the
interaction between neutron spin and the Earth's rotation by using
neutron interferometry experiment\cite{Anandan,Overhauser}, which
has been successful in investigating the Aharonov-Carmi effect,
the gravitational analogue to the Aharonov-Bohm effect in
electrodynamics.

\end{document}